\documentclass{ws-mpla}

\begin{document}

\markboth{Merab Gogberashvili and Ramaz Khomeriki}
{Trapping of Nonlinear Gravitational Waves by Two-Fluid Systems}

\catchline{}{}{}{}{}

\title{Trapping of Nonlinear Gravitational Waves by Two-Fluid Systems}

\author{\footnotesize Merab GOGBERASHVILI}

\address{Andronikashvili Institute of Physics,
6 Tamarashvili Street, Tbilisi 0177, Georgia \\
Javakhishvili Tbilisi State University,
3 Chavchavadze Avenue, Tbilisi 0128, Georgia \\
gogber@gmail.com}

\author{Ramaz KHOMERIKI}

\address{Javakhishvili Tbilisi State University,
3 Chavchavadze Avenue, Tbilisi 0128, Georgia\\
 Max-Planck-Institut f\"{u}r Physik komplexer Systeme,
01187 Dresden, Germany\\
khomeriki@hotmail.com
}

\maketitle


\begin{abstract}
We show that the coupled two-fluid gravitating system (e.g. stiff matter and 'vacuum energy') could trap nonlinear gravitational waves (e.g. Einstein-Rosen waves). The gravitational wave amplitude varies harmonically in time transferring the energy coherently to the stiff matter wave, and then the process goes to the backward direction. This process mimics the behaviour of trapped electromagnetic waves in two-level media. We have defined the limits for the frequency of this energy transfer oscillations.

\end{abstract}

\ccode{04.30.-w, 47.35.Bb, 98.80.Cq} 

\medskip


Investigation of the nonlinearly coupled state of electromagnetic waves with two or three level medium has become one of the hottest research spots this last decade\cite{LAB1,LAB2,LAB3}. The most exciting feature of this nonlinear effect is that the electromagnetic wave, which has a fixed propagation velocity in vacuum, could form with medium some breather-like structures localized in space. Such systems allows not only to slow down, and eventually stop\cite{XZP} a light pulse\cite{KKA1,KKA2,KKA3}, but also there exists a possibility to release it, thus creating a {\it gap soliton memory}\cite{Me-Ai}. In the simplest case of two-level medium the governing equations are Maxwell-Bloch system\cite{AP1,AP2} and, as the very recent theoretical investigations show\cite{Kh-Le}, one can control the mobility of the light pulse by the stationary radiation with a different polarization.

We want to find out if similar effect exists for nonlinear gravitational waves inside of gravitating multi-component fluid systems, which are analogous to multi-level mediums in nonlinear optics. 

Existence of gravitational waves is generally accepted, at least since their indirect observation via energy and angular momentum losses in binary pulsars (see e.g.\cite{Pulsar} and references therein). The study of gravitational waves involves mainly the weak field approximation of Einstein's equations (in close analogy with what is normally done in electromagnetism), which is obviously the relevant regime for existing gravitational wave detectors. On the other hand, the non linearity of the gravitational field is one of its most characteristic properties. It is likely that some of the crucial properties of the gravitational field show themselves only through the nonlinear terms. Thus, a number of exact nonlinear gravitational wave solutions have been found\cite{Exact,Bicak}.

In\cite{Ste} it was shown that standing gravitational waves do not exist for the main classes of exact, wave-like solutions to the Einstein equations. The only exception is the Einstein-Rosen solution, which will be considered in this paper. The exact nonlinear plane-symmetric standing gravitational wave solution, but bounded with domain wall, was found in\cite{GMS}. 

Einstein-Rosen cylindrical waves are among the simplest non-stationary solutions to the vacuum Einstein equations\cite{Exact,Bicak}. Despite the fact that cylindrically symmetric waves cannot describe exactly the radiation from bounded sources, they have played an important role in many physical models\cite{Bicak}. Also we note that cylindrical gravitational waves are related to a special class of spherical and toroidal waves\cite{Mar1,Mar2}.

One of the important features of the Einstein-Rosen wave is that it is special case of the gravitational Bessel beam\cite{Kram}. Bessel beams are cylindrically symmetric solutions to the wave equations, which were extensively studied and applied in both optics and acoustics (see the reviews\cite{Bessel1,Bessel2,Bessel3}). The peculiarity of this type of wave is that it is well localized in space. This is the property we are looking for in interactions of gravitational waves with media.
 
There have been numerous investigations on the interaction between multi-component plasma and linear gravity waves (see\cite{plasma1,plasma2,plasma3} and references therein). In this paper we consider the nonlinear case. For the simplicity we assume coexistence of two perfect fluids with zero relative velocity that have their own energy momentum tensors of the form:
\begin{equation} \label{T}
T^i_{\mu\nu} = (\epsilon^i + p^i) u_\mu u_\nu - g_{\mu\nu}p^i~,
~~~~~ i = 1,2
\end{equation}
where $\epsilon^i$ are the energy densities, $p^i$ are the pressures and $u^\nu$ is the normalized 4-velocity vector,
\begin{equation}
u^\mu u_\mu = 1~.
\end{equation}

As one of the ingredients of our two-fluid system we take the negative vacuum energy,
\begin{equation}
\epsilon^1 = - p^1 = - \rho~,
\end{equation}
with the energy-momentum tensor
\begin{equation}
T^1_{\mu\nu} = - g_{\mu\nu}\rho~,
\end{equation}
which can be associated to one or more scalar fields\cite{Pe-Ra}. Having in mind the electromagnetic analogy we assume that under the influence of nonlinear gravitational wave the vacuum energy $\rho$ can oscillate around it's average value (but stay always positive).

For the second fluid we denote $\epsilon^2 = \epsilon$, $p^2 = p$ and consider the equation of state
\begin{equation} \label{e=p}
\epsilon = p + 2\rho ~,
\end{equation}
which for the case of $\rho \rightarrow 0$ reduces to the equation of state of the stiff matter,
\begin{equation}
\epsilon = p~.
\end{equation}
The study of stiff fluid in gravitational waves contexts is interesting in the sense that in such matter the speed of sound is equal to the speed of light\cite{Zel1,Zel2}. The equation of state (\ref{e=p}) corresponds to almost maximum stiffness compatible with causality and perhaps is good approximation for supernuclear densities of matter existing at early stages of the universe expansion and in the centers of very massive stellar objects. Note that since we assume $\rho > 0$, then $\epsilon \geq p$ and the equation of state for the second fluid (\ref{e=p}) is below the stiffness limit.

The gravitational equations for our two-fluid system
\begin{equation} \label{Einstain}
R_{\mu\nu} - \frac 12 g_{^\mu\nu} R = k \left( T^1_{\mu\nu} +
T^2_{\mu\nu}\right)~,
\end{equation}
where $k = 8\pi G/c^4$ is the Einstein constant, is equivalent to the Einstein equations in anti de Sitter space (with the negative cosmological constant $- k\rho$). Problems of finding exact wave solutions of (\ref{Einstain}) within matter are in general very complicated, since one has to deal with interaction between the gravitational waves and the sound waves excited by them. Even for the empty space-time there exists only few exact wave solutions\cite{Exact,Bicak}. As it was mentioned above, in this paper we consider the most studied wave-like solution, which is based on  the cylindrically symmetric Einstein-Rosen {\it ansatz}\cite{EW1,EW2}
\begin{equation}\label{Ansatz}
ds^2 = e^{2(\gamma - \psi)}\left( dt^2 - dr^2\right) - r^2
e^{-2\psi} d\theta^2 - e^{2\psi}dz^2~,
\end{equation}
where the quantities $\gamma$ and $\psi$ are functions of the time $t$ and the radial coordinate $r$ only.

As was shown in\cite{Ta-Ta} for any perfect fluid with a non-zero pressure, the energy-momentum tensor and its conservation can be written in terms of a scalar function. The non-rotating stiff fluid obeys particularly simple equations that are equivalent to those of the massless scalar field $\sigma$,
\begin{equation} \label{sigma}
\frac{1}{\sqrt{-g}}~\partial_\mu (\sqrt{-g} g^{\mu\nu}\partial_\nu
\sigma) = \Box\sigma = \ddot \sigma - \sigma'' - \frac 1r \sigma' = 0~,
\end{equation}
where overdots mean time derivatives and primes stand for derivatives with respect of radial coordinate. 

Now we want to modify the representation used in\cite{Ta-Ta} for stiff matter to accommodate it for our two fluid system. The simplest possibility is to shift $\epsilon$ and $p$ by $\rho$ and describe the fluid characteristics by the scalar function $\sigma$ in the form:
\begin{equation}
\epsilon - \rho = p + \rho = \frac {1}{k} ~\partial_\nu \sigma
\partial^\nu \sigma ~, ~~~~~
u^\mu = \frac{\partial^\mu
\sigma}{\sqrt{\partial^\nu \sigma\partial_\nu \sigma}}~.
\end{equation}
The right hand side of the Einstein equations (\ref{Einstain}) becomes equivalent to the energy-momentum tensor of scalar field,
\begin{equation}
k ~(2 u_\mu u_\nu - g_{\mu\nu}) (p +\rho )= 2~\partial_\mu
\sigma\partial_\nu \sigma - g_{\mu\nu} \partial^\alpha
\sigma\partial_\alpha \sigma~.
\end{equation}
So (\ref{Einstain}) may be written in the simple form:
\begin{equation} \label{R=sigma}
R_{\mu\nu} = 2 \partial_\mu \sigma\partial_\nu \sigma~.
\end{equation}
For fluids that are not stiff the equations analogous to (\ref{sigma}) and (\ref{R=sigma}) are more complicated and not amenable to the separation of variables employed in this paper.

Non-zero components of the Ricci tensor for the {\it ansatz} (\ref{Ansatz}) are\cite{EW1,EW2}
\begin{eqnarray}
R_{tt} &=& \ddot \psi - \psi'' - \psi'/r - 2 \dot \psi ^2 -
\ddot \gamma + \gamma'' + \gamma'/r ~, \nonumber \\
R_{rr} &=& -\ddot \psi + \psi'' + \psi'/r - 2 \psi'^2 +
\ddot \gamma - \gamma'' + \gamma'/r~, \nonumber \\
R_{\theta\theta} &=& - r^2 e^{-2\gamma}
\left(\ddot \psi - \psi'' - \psi'/r \right)~, \\
R_{zz} &=& e^{-2\gamma + 4\psi}
\left(\ddot \psi - \psi'' - \psi'/r\right)~, \nonumber \\
R_{rt} &=& - 2 \dot\psi\psi' + \dot\gamma /r~. \nonumber
\end{eqnarray}
Then the Einstein equations in our case (\ref{R=sigma}) reduce to
\begin{eqnarray} \label{Le-Te}
\Box\sigma = \Box\psi = 0~, \nonumber \\
\dot\gamma = 2 r \left(\dot\psi\psi' + \dot\sigma\sigma'\right)~,
\nonumber \\
\gamma' = r \left( \dot\psi^2 + \psi'^2 + \dot\sigma^2 +
\sigma'^2 \right)~, \\
\epsilon - \rho = p + \rho = \frac {1}{k}~\left(\dot\sigma^2 -
\sigma'^2\right)e^{-2(\gamma - \psi)}~.\nonumber
\end{eqnarray}

We see that both the fields, the fluid factor $\sigma$ and the metric component $\psi$, behave like massless, minimally coupled, source-free scalar fields. Finding $\sigma$ and $\psi$ from the first equations of the system (\ref{Le-Te}) one can find expression of the second unknown metric parameter $\gamma$ in the form\cite{Le-Ta}:
\begin{equation} \label{gamma}
\gamma = \int dr ~r (\dot \sigma^2 + \sigma'^2 + \dot \psi^2 +
\psi'^2)  +\int dt ~ 2r(\dot\sigma\sigma' + \dot
\psi \psi')~.
\end{equation}

Some well studied special cases of the system (\ref{Le-Te}) and (\ref{gamma}) are:

$a).$ If we assume that the vacuum energy is equal to zero, $\rho=0$, and also choose the solutions of (\ref{Le-Te}) in the form:
\begin{equation}
\sigma = a t~, ~~~~~ \psi = 0~,
\end{equation}
where $a$ is a constant, we get singular-free, static, cylindrically symmetric solution of Einstein's equations for stiff matter\cite{Kra},
\begin{eqnarray}\label{Kramer}
ds^2 = e^{a^2 r^2}\left(dt^2 - dr^2\right) - r^2d\theta^2 - dz^2~,
\nonumber \\
\epsilon = p = \frac {a^2}{k}e^{-a^2 r^2}~.
\end{eqnarray}

$b).$ If again we suppose that  $\rho =0$ and take
\begin{equation}
\sigma = 0~,
\end{equation}
the Einstein equations (\ref{Einstain}) reduce to another known system of equations describing cylindrical gravitational waves in empty space ($\epsilon = p = 0$)\cite{EW1,EW2},
\begin{eqnarray} \label{Ei-Ro}
\ddot \psi - \psi'' - \frac 1r \psi' = 0~, \nonumber \\
\dot\gamma = 2 r \dot\psi\psi'~, \\
\gamma' = r \left( \dot\psi^2 + \psi'^2 \right)~. \nonumber
\end{eqnarray}

Now let us consider the case that we are looking for. Suppose the core of some supermassive object obeys the equation of state (\ref{e=p}). The scalar modes of the stiff matter fluctuations $\sigma$ are indistinguishable from the excitations of metric $\psi$. In this resonant case we write,
\begin{equation}
\psi (t,r) = \sigma (t,r) ~,
\end{equation}
and the main system (\ref{Le-Te}) again reduces\cite{Let} to the equations for cylindrical gravitational waves (\ref{Ei-Ro}), but waves are now inside our two-component medium ($\epsilon, p, \rho \ne 0$). So we have cylindrical gravitational waves inside of almost stiff matter with the energy density
\begin{equation}
\epsilon  = p + 2 \rho = \rho + \frac {1}{k}(\dot\psi^2 -
\psi'^2)e^{-2(\gamma - \psi)}~.
\end{equation}

Now we approaching the main task of our paper, finding stationary coupled matter-gravitational wave solutions of the system (\ref{Le-Te}),
\begin{equation} \label{sigma-psi1}
\psi (r,t) = f(r)\sin(\omega t)~, ~~~~~
\sigma (r,t)=
f(r)\cos(\omega t)~,
\end{equation}
where, according to  the first line of (\ref{Le-Te}), $f(r)$ satisfies the equation
\begin{equation} \label{sigma-psi2}
f''+ \frac 1r f' + \omega^2 f= 0~.
\end{equation}
For (\ref{sigma-psi1}) the second equation of (\ref{Le-Te}) gives the trivial relation:
\begin{equation} \label{sigma-psi3}
\dot\gamma =0~.
\end{equation}
Thus $\gamma\equiv\gamma(r)$ is a function of the radial coordinate $r$ only. Substituting (\ref{sigma-psi1}) into the third equation of (\ref{Le-Te}) we get:
\begin{equation} \label{sigma-psi4}
\gamma' = r \left(\omega^2f^2+ f'^2\right)~.
\end{equation}
Putting now (\ref{sigma-psi1}) into the fourth equation of (\ref{Le-Te}) one finds:
\begin{equation} \label{sigma-psi5}
\epsilon = p + 2\rho =\rho + \frac
{1}{k}~\left[\omega^2\sin^2(\omega t)f^2- \cos^2(\omega t)f'^2\right]e^{-2[\gamma(r) - \psi(r,t)]}~,
\end{equation}
where  $\psi(r,t)$ and $\gamma(r)$ are defined via $f(r)$ according to the relations (\ref{sigma-psi1}) and (\ref{sigma-psi4}), respectively.

Note that the simplest solution of (\ref{sigma-psi2}) is a zero order Bessel function of the first kind,
\begin{equation}
f(r) = J_0(\omega r)~,
\end{equation}
which has the following asymptotics (see e.g.\cite{book}):
\begin{eqnarray} \label{sigma-psi8}
r &\rightarrow& 0   ~~~~~~ f(r) \approx 1-\frac{\omega^2r^2}{4}; \nonumber \\
r &\rightarrow& \infty ~~~~ f(r) \approx \sqrt{\frac{2}{\pi \omega
r}}\cos\left(\omega r-\frac{\pi}{4}\right).
\end{eqnarray}
Thus at the infinity ($r \rightarrow \infty$),
\begin{equation}
\gamma \rightarrow \omega r~, ~~~~~
\epsilon \rightarrow \rho + e^{-\omega r}~,
\end{equation}
and the energy $\epsilon$ decays exponentially to the 'vacuum energy' value $\rho$. So (\ref{sigma-psi1}) is a particular case ($z$-independent solution) of Bessel waves extensively studied in recent years\cite{Bessel1,Bessel2,Bessel3}. 

The coherent process described by the solution (\ref{sigma-psi1}) mimics very much the behaviour of electromagnetic waves in two-level media\cite{LAB1,LAB2,LAB3,XZP,KKA1,KKA2,KKA3,Me-Ai,Kh-Le}, where the electromagnetic waves are periodically absorbed and radiated by the two-level atoms, and thus the electromagnetic waves are trapped all the time inside the medium. Similarly, in our case gravitational waves are trapped by two-fluid system transferring periodically the energy to the matter waves and then this energy restores again.

According to the requirement that the energy in (\ref{sigma-psi5}) should be positive, it follows the restriction on the oscillation
frequency
\begin{equation}
\omega<\Omega\simeq\sqrt{k\rho}~.
\end{equation}
This itself means that the radial size of the object that traps nonlinear gravitational wave, as it follows from the same relation (\ref{sigma-psi5}), should be larger than the characteristic length
\begin{equation} \label{l}
l\simeq \frac 1\Omega=\frac 1{\sqrt{k\rho}}~.
\end{equation}

Supposing that $k\rho$ is the ordinary cosmological constant we find unacceptably large the value of the parameter (\ref{l}) (of order of the horizon $10^{28} cm$), as astronomical observations imply\cite{lambda1,lambda2,lambda3} the present value of cosmological constant cannot exceed $10^{-56} cm^{-2}$. However, we introduce the fluid with the properties of the cosmological constant as one of the ingredients of our two component system only for the simplicity of considerations. Vacuum energy is just needed as reference to keep positive the energy of the system during coherent oscillations of gravitational and matter waves. Also problems of the definition of gravitational energy (in general) and equations of states for very dense objects are well known. The controversial point about the concept of energy localization\cite{energy1,energy2,energy3} is the origin of a long-standing discussion on the energy and momentum carried by a gravitational wave. Perhaps one can find other two (or more) component gravitating systems that will provide trapping of nonlinear gravitational waves of different kind by the mechanism considered in this article.

Concluding we have investigated the coupled stationary state of a two-fluid gravitating system (stiff matter plus vacuum energy) with trapped Einstein-Rosen type nonlinear gravitational waves. In this medium the gravitational wave amplitude varies harmonically in time transferring the energy coherently to the stiff matter wave, and then the process goes to the backward direction, i.e. matter waves transfer the energy to the gravitational waves. We have defined the limits for the frequency of this energy transfer oscillations.


\section*{Acknowledgements}

M. G. acknowledges the hospitality extended during his visits at the Abdus Salam International Centre for Theoretical Physics where
this paper was prepared.

R. Kh. acknowledges financial support of the Georgian National Science Foundation (Grant No GNSF/STO7/4-197) and USA Civilian Research and Development Foundation (award No GEP2-2848-TB-06).


\end{document}